# TOWARDS CLASSIFYING BENIGN AND MALICIOUS PACKAGES USING MACHINE LEARNING

**Thanh-Cong Nguyen [1], Ngoc-Thanh Nguyen [2], Van-Giau Ung [3], Duc-Ly Vu [4] (*)**

[1] *University of Information Technology,*
[2,3,4] *School of Computing and Information Technology;*
[1] *20521143@gm.uit.edu.vn,* [2] *thanh.nguyenngoc,* [3] *giau.ung,* [4] *ly.vu@eiu.edu.vn*

(*) Corresponding author

***Abstract***: Recently, the number of malicious open-source packages in package repositories has been increasing dramatically. While major security scanners focus on identifying known Common Vulnerabilities and Exposures (CVEs) in open-source packages, there are very few studies on detecting malicious packages. Malicious open-source package detection typically requires static, dynamic analysis, or both. Dynamic analysis is more effective as it can expose a package's behaviors at runtime. However, current dynamic analysis tools (e.g., *ossf's package-analysis*) lack an automatic method to differentiate malicious packages from benign packages. In this paper, we propose an approach to extract the features from dynamic analysis (e.g., executed commands) and leverage machine learning techniques to automatically classify packages as benign or malicious. Our evaluation of nearly 2000 packages on npm shows that the machine learning classifier achieves an AUC of 0.91 with a false positive rate of nearly 0%.

## I. INTRODUCTION

Modern software development involves extensive use of open-source packages. Using third-party packages not only saves development costs but also improves efficiency in software development. While open-source packages offer significant benefits, they also pose security risks in both the packages themselves and the repositories they are hosted (Ladisa et al., 2023). For example, attackers can insert malicious code into upstream open-source libraries, compromising end users. Several malicious actions on the victim's computer include stealing information (operating system and user information, tokens, session keys, etc.), executing malicious commands or initiating a reverse shell. (Guo et al., 2023).

Several malware detection techniques have been proposed by academic researchers and industries (Ladisa et al., 2023) to flag malicious open-source packages. Generally, there are two main techniques: static and dynamic analysis. Static analysis techniques only examine the package's metadata or source code without executing it, while dynamic analysis techniques execute the package's code, typically in an isolated environment (e.g., in a sandbox). Although static analysis techniques are fast and easy to implement, they are ineffective in dealing with obfuscated packages (Moser et al., 2007) and may introduce many false positives (Vu et al., 2023). Dynamic analysis techniques are considered to be more precise and reliable as they expose the package's behaviors at runtime. For this reason, in this paper, we focus on dynamic analysis of open-source packages to observe their actual behavior, in particular, those that extract information from the execution traces and network traffic.

There are a few dynamic analysis techniques/tools in the wild. The tool *package-analysis*, developed by the Open Source Security Foundation (OpenSSF) is one of the most popular and actively maintained dynamic analysis tools. *package-analysis* looks for behaviors that indicate malicious software: *1) What files do they access? 2) What addresses do they connect to? and 3) What commands do they run?*. *package-analysis* captures malicious interactions with the system as well as network connections that could be used to leak sensitive data or allow remote access using Gvisor sandbox (gVisor). In addition, the raw outputs of *package-analysis*s are available on Google BigQuery (OSSF) enabling us to perform in-depth analysis of common behaviors of benign and malicious open-source packages.

However, the *package-analysis* tool only provides raw analysis results of the package's behaviors (in JSON format) that require additional effort to determine whether a package is malicious. Hence, our work contributes to filling this gap by processing the package's raw analysis, extracting suspicious features of the packages, and applying machine learning techniques to classify packages as benign or malicious. In summary, this paper makes the following contributions:

- We investigate a set of features to characterize malicious open-source packages based on dynamic analysis.
- We propose a machine learning-based approach to classify benign and malicious packages using the proposed features extracted from dynamic analysis.
- We evaluate the performance of different machine learning-based classification models through 10-fold cross-validation method.

## II. METHODOLOGY

### A. Data Collection

Table 1 shows the number of benign and malicious packages in our dataset. The samples in our dataset are collected from real-world attacks (for the malicious packages set) and from official package repositories (for the benign packages set).

**Table 1.** Our dataset

| Categories | Repository | No. of packages |
|---|---|---|
| Malicious | npm | 1170 |
| Benign | npm | 999 |

#### a) Malicious packages

Figure 1 illustrates our data collection and analysis process. We query the following online sources to collect malicious open-source packages:

- *Vulert* (Vulert): *Vulert* provides security advisories about open-source packages in popular package repositories such as npm, PyPI, RubyGems, and Crates.io.
- *Vulners (Inc.)*: *Vulners* provides a database of security vulnerabilities in open-source packages
- *OSV (OSV)*: *OSV* is a distributed vulnerability database for open-source packages maintained by Google.

These services provide free APIs to search for and download known malicious packages from package repositories such as PyPI, npm, Crates.io, etc. We also use security services such as VirusTotal (VirusTotal) to label a package as benign or malicious based on the package's features (Section II.C). For example, a package is malicious if it contacts a domain flagged as malicious by VirusTotal or executes sensitive commands such as *curl* or *exec* Linux commands.

In our dataset, npm contains the majority of malicious packages. Our data curation includes 113 PyPI packages, 1041 npm packages, 16 RubyGems packages, and one crates.io package. Since machine learning techniques require a sufficient number of training examples, we chose npm packages instead of other types of packages. It should be noted that our results (Section III) may not generalize to packages from other repositories, such as PyPI or RubyGems..

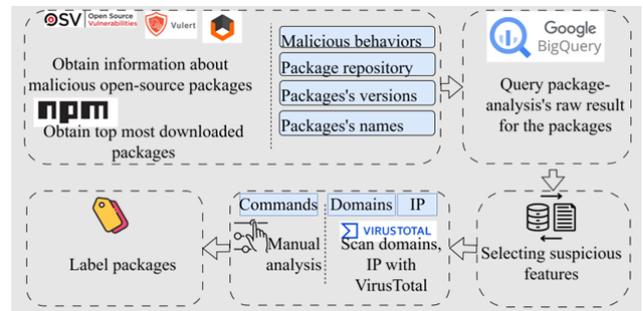

**Figure 1.** Data Collection and Analysis Workflow.

#### b) Benign packages

Following the approaches in (Vu et al., 2023) and (Zahan et al., 2022), we select the top 1000 popular packages on npm as benign packages. We chose npm packages because they are the most common packages available in the online package-analysis's BigQuery dataset (see Section II.A). After collecting the list of popular npm packages, we queried their analysis results in the package-analysis dataset on Google BigQuery (OSSF).

### B. Data preprocessing

Since the raw package-analysis results are published in a database of tables, we first convert them into CSV format, which is more user-friendly for machine learning algorithms. Then, we remove unnecessary fields and duplicates in the CSV files. We also exclude rows that have missing data.

### C. Feature Engineering

Our exploratory analysis, utilizing visualization and statistical methods on the raw package analysis results of the packages in Table 1, identified the following features:

- **Executed commands:** Number of executed commands performed by a package.
- **Domains:** Number of domains connected by a package.
- **IP addresses:** Number of IP addresses contacted by a package.

All these features are of integer data type. We employ one-hot encoding to transform the features into numerical vectors, which are subsequently fed directly into machine learning models. After that, the feature vectors are fed directly into machine learning models.

*D. Model Evaluation*

In this step, we select 17 machine learning models available in the scikit-learn framework (Pedregosa et al., 2011). The performance of these models is evaluated using the following metrics:

- **Accuracy:** this metric indicates the frequency of correct predictions of ML models.
- **Precision:** this metric shows the frequency of correct predictions of ML models for positive samples.
- **Recall:** this metric shows the frequency with which ML models can detect true positive samples correctly from all positive samples in our dataset.
- **F1 Score:** this metric is the harmonic mean of Precision and Recall.
- **False Positive Rate (FPR):** this metric measures the proportion of negative samples identified as positive.
- **False Negative Rate (FNR):** this metric measures the proportion of positive samples identified as negative.
- **Receiver Operating Characteristic Curve (ROC Curve)**: this graph illustrates a classification models' performance across all thresholds. This graph represents two parameters: True Positive Rate (TPR) and False Positive Rate (FPR).
- **Area under the ROC Curve (AUC):** this metric measures the two-dimensional area beneath the ROC Curve, which means the distinguishing ability of classification models.

In the evaluation phase, we employ 10-fold cross-validation, which first randomly divides all the data into ten parts, then holds out 10% of the data for testing (Kohavi, 1995). This process is repeated ten times, after which the mean accuracy for the algorithm is calculated. Tables 2 and 3 in this paper report the performance of each machine learning model using 10-fold cross-validation.

## III. RESULTS

Figure 1 shows the Receiver Operating Characteristic (ROC) curves (ROC) for all models. Overall, the curves lean towards the top left corner, indicating that our predictive models are highly accurate in classifying benign and malicious open-source packages. Tree-based classifiers outperform the other models, especially when boosting techniques are applied. Notably, three of the top-performing machine learning models in Tables 1 and 2 are tree-based. As shown in Figure 1, the Logistic Regression and Gaussian-based classifiers have the poorest performance among the evaluated models.

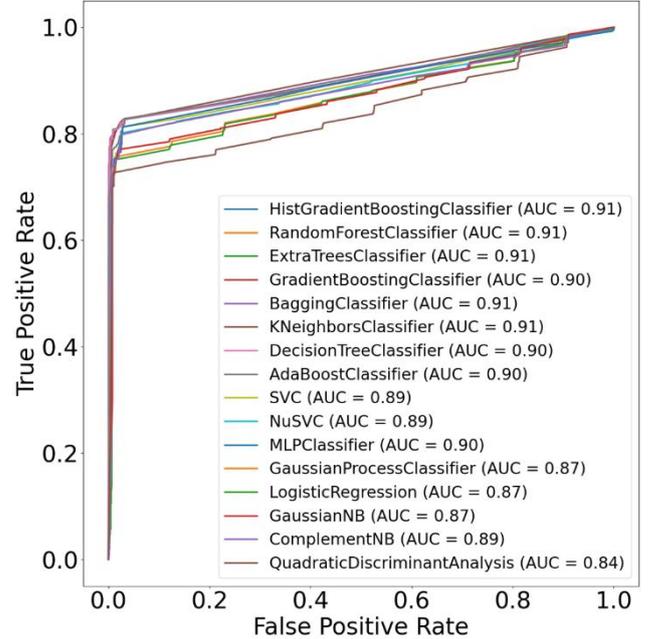

**Figure 2.** ROC Curves of the Machine Learning Models.

Tables 2 and 3 present the top machine learning models sorted by AUC. The models perform well on both training and testing sets across all evaluation metrics. This suggests that the models do not suffer from overfitting. For example, the difference between the median AUC of all models in the training and testing phases is 0.004, which is relatively small.

However, the false negative rates are slightly higher than the false positive rates, indicating that the models sometimes fail to detect malicious packages. This implies that the models require more information about the packages to classify benign and malicious packages more accurately. Additionally, the average accuracy of the models is 0.923, which aligns closely with the expectations of package repository maintainers (Vu et al., 2023).

The accuracy of the models may not be practically optimal, as shown in Table 1 (an average of 0.8935) and Table 2 (an average of 0.8898). However, the precision and recall values of the models in the training phase, as shown in Table 1, are above 90%. These results are promising and suggest that even with a relatively small sample size, it is possible to build effective predictive models for classifying malicious and benign packages.

**Table 2.** Performance of the Top Machine Learning Models on the Training Set.

| Model | Accuracy | Precision | Recall | F1 | FPR | FNR | AUC |
|---|---|---|---|---|---|---|---|
| DecisionTreeClassifier | 0.8940 | 0.9005 | 0.9015 | 0.8940 | 0.0199 | 0.1772 | 0.9116 |
| ExtraTreesClassifier | 0.8940 | 0.9005 | 0.9015 | 0.8940 | 0.0199 | 0.1772 | 0.9116 |
| HistGradientBoostingClassifier | 0.8933 | 0.8994 | 0.9006 | 0.8933 | 0.0226 | 0.1762 | 0.9112 |
| BaggingClassifier | 0.8937 | 0.8999 | 0.9011 | 0.8937 | 0.0216 | 0.1762 | 0.9111 |
| RandomForestClassifier | 0.8940 | 0.9003 | 0.9014 | 0.8940 | 0.0207 | 0.1765 | 0.9111 |
| GradientBoostingClassifier | 0.8934 | 0.8990 | 0.9005 | 0.8934 | 0.0247 | 0.1744 | 0.9110 |
| KNeighborsClassifier | 0.8922 | 0.8988 | 0.8998 | 0.8922 | 0.0214 | 0.1791 | 0.9092 |

**Table 3.** Performance of the Top Machine Learning Models on the Validation Set.

| Model | Accuracy | Precision | Recall | F1 | FPR | FNR | AUC |
|---|---|---|---|---|---|---|---|
| HistGradientBoostingClassifier | 0.8897 | 0.8961 | 0.8966 | 0.8894 | 0.0263 | 0.1805 | 0.9103 |
| RandomForestClassifier | 0.8907 | 0.8968 | 0.8975 | 0.8904 | 0.0263 | 0.1787 | 0.9102 |
| ExtraTreesClassifier | 0.8904 | 0.8968 | 0.8973 | 0.8901 | 0.0258 | 0.1797 | 0.9099 |
| GradientBoostingClassifier | 0.8892 | 0.8950 | 0.8958 | 0.8889 | 0.0297 | 0.1786 | 0.9092 |
| BaggingClassifier | 0.8882 | 0.8946 | 0.8950 | 0.8879 | 0.0281 | 0.1819 | 0.9087 |
| KNeighborsClassifier | 0.8909 | 0.8981 | 0.8980 | 0.8906 | 0.0221 | 0.1819 | 0.9081 |
| DecisionTreeClassifier | 0.8894 | 0.8960 | 0.8964 | 0.8891 | 0.0258 | 0.1814 | 0.9081 |

## IV. THREATS TO VALIDITY

This section discusses the factors that could have influenced our work.

*We considered only the top 1000 npm packages* as benign, out of more than two million packages (npm). A larger number of packages would be necessary for a comprehensive ecosystem analysis and for training machine learning models.

*The machine learning models focus on the Javascript packages in npm, particularly JavaScript files.* Extending this approach to other interpreted languages and file types appears straightforward (e.g., Python/PyPI and Ruby/RubyGems) by collecting more samples from Python package repository and train the models on the samples.

*The malicious dataset used in our study may not accurately represent malicious packages in the wild.* Not all malicious npm packages are publicly known. *Vulert, Vulners,* and *OSV* are the largest repositories for malicious packages available for researchers.

*We rely on package-analysis to extract the features of a package.* In our observation (Nguyen et al., 2024), *package-analysis* is ineffective at analyzing packages at the installation phase. This limitation prevents us from capturing the desired behavior of open-source packages.

*The dynamic analysis tool package-analysis currently operates on Linux operating systems only.* This limitation arises because its sandbox supports only Linux. Future work will involve extending package-analysis's sandbox to support other operating systems, such as Windows and macOS by modifying the sandbox to support other file formats such as Portable Executable for Windows.

## V. CONCLUSION AND FUTURE WORK

In this paper, we propose a machine learning approach to classify benign and malicious packages. Our approach relies on features extracted through dynamic analysis, including executed commands, IP addresses, and domains. We applied 17 machine learning models available in *scikit-learn* to a dataset of benign and malicious packages. Our evaluation shows that the models performed relatively well across all metrics, particularly in terms of false positive rates.

Our next step is to investigate additional package features, such as those derived from static analysis, to improve the performance of the machine learning models Additionally, we plan to extend the evaluation to packages from other ecosystems, such as PyPI and RubyGems. This will require additional effort in curating packages, particularly malicious ones.

To deploy our solution in practice, we plan to integrate our machine learning models into existing package repositories, such as PyPI, or provide a third-party tool to support the detection of malicious code in open-source packages. This integration will require developing a new malware detection tool that can efficiently and promptly scan open-source packages.

To promote openness in science, we have made the scripts and results of our evaluation publicly available at https://github.com/ngoiThu0/thesis.